# Pseudo-Haptic Button for Improving User Experience of Mid-Air Interaction in VR

Woojoo Kim, Shuping Xiong[*]

Human Factors and Ergonomics Laboratory, Department of Industrial and Systems Engineering, College of Engineering, Korea Advanced Institute of Science and Technology (KAIST), 291 Daehak-ro, Yuseong-gu, Daejeon 34141, Republic of Korea; {xml1324, shupingx}@kaist.ac.kr

[*] Corresponding author (Prof. Shuping Xiong), Telephone: +82-42-350-3132, Fax: +82-42-350-3110, Email: shupingx@kaist.ac.kr

**Abstract.** Mid-air interaction is one of the promising interaction modalities in virtual reality (VR) due to its merits in naturalness and intuitiveness, but the interaction suffers from the lack of haptic feedback as no force or vibrotactile feedback can be provided in mid-air. As a breakthrough to compensate for this insufficiency, the application of pseudo-haptic features which create the visuo-haptic illusion without actual physical haptic stimulus can be explored. Therefore, this study aimed to investigate the effect of four pseudo-haptic features: proximity feedback, protrusion, hit effect, and penetration blocking on user experience for free-hand mid-air button interaction in VR. We conducted a user study on 21 young subjects to collect user ratings on various aspects of user experience while users were freely interacting with 16 buttons with different combinations of four features. Results indicated that all investigated features significantly improved user experience in terms of haptic illusion, embodiment, sense of reality, spatiotemporal perception, satisfaction, and hedonic quality. In addition, protrusion and hit effect were more beneficial in comparison with the other two features. It is recommended to utilize the four proposed pseudo-haptic features in 3D user interfaces (UIs) to make users feel more pleased and amused, but caution is needed when using proximity feedback together with other features. The findings of this study could be helpful for VR developers and UI designers in providing better interactive buttons in the 3D interfaces.

## 1. Introduction

Virtual Reality (VR) is a concept that has existed for several decades but has not gained much attention from the general public until the recent advancement of mass-market VR head-mounted displays (HMDs), which have renewed the interest in the design of 3D user interfaces (UIs) and 3D interaction techniques in immersive virtual environments. In such environments, head and hand tracking are often enabled by 6 degree-of-freedom tracking technologies that potentially provide natural and direct interaction with objects stereoscopically displayed in the virtual world. Consequently, the user can manipulate virtual objects within reach in a similar way to grabbing or selecting objects in the real world.

Selection is one of the fundamental tasks of user interaction in VR and 3D UIs (LaViola Jr. et al., 2017), and button is one of the fundamental widgets in UI both physically and digitally (Janlert, 2014), inevitably evolving to the 3D form in VR (Speicher et al., 2019; van Dam, 1997). However, current 3D UIs are usually designed by adapting directly from the paradigm of 2D UIs (Lee et al., 2018) which leads to a gap in the user experience (Bowman et al., 2012). Interacting with the 2D UI in 3D space is known to be demanding due to the inherent difficulty of understanding and performing actions in 3D space (Herndon et al., 1994).

In the HMD-based VR, head-gaze, hand-held controllers, and free-hand mid-air gestural inputs are considered typical interaction modalities (Speicher et al., 2018). Although the controller-based interaction turned out to be more responsive than the free-hand mid-air interaction (Caggianese et al., 2019; Dudley et al., 2019), mid-air interaction is regarded as a natural and intuitive way of interacting with 3D UIs in VR (Lee and Hui, 2018). Since the release of the Leap Motion controller which supports real-time skeletal tracking of the user's hands and fingers, steady efforts on technological advancements have been driven in the field of vision-based hand tracking. The current market-dominating VR and augmented reality (AR) HMDs such as Oculus Quest 2 and Microsoft Hololens 2 are coming out with a decent real-time hand tracking capability enabled by inputs from embedded monochrome (Han et al., 2020) or depth (Ungureanu et al., 2020) cameras, facilitating a simple and easy implementation of controller-free mid-air interactions in VR.

Although mid-air interaction in VR demonstrates the remarkable advantages of naturalness and intuitiveness, the usability of VR UIs is highly dependent on the availability of haptic feedback. Many studies have proven that the interaction suffers when no haptic feedback is provided (Benko et al., 2006; Dudley et al., 2019; Faeth and Harding, 2014; Hoggan et al., 2008; Shneiderman, 1997; Zhao et al., 2014). In order to provide the haptic feedback to inform users about the confirmation of their action, studies have used active haptics with wearable haptic devices such as smart gloves/thimbles (Blake and Gurocak, 2009; Bouzit et al., 2002; Bullion and Gurocak, 2009; Gabardi et al., 2016; Kim et al., 2016) and contactless modalities such as ultrasounds (Arafsha et al., 2015; Carter et al., 2013; Freeman et al., 2019; Shakeri et al., 2018), while passive haptics has been enabled by physical props (Azmandian et al., 2016; Joyce and Robinson, 2017; Strandholt et al., 2020). However, such approaches for providing the haptic feedback would eventually require compatible equipment, resulting in ineffective and bulky experience in real-life VR usages (Hwang et al., 2017), and not always available to users taking the practical considerations into account. Therefore, the lack of haptic feedback remains as the common problem of current mid-air button interaction.

The pseudo-haptic method which uses multi-sensory contradiction, where visual cues create haptic illusions without actual physical haptic stimulus (Hachisu et al., 2011) can be a breakthrough for compensating this issue (Chattopadhyay, 2016). Although it has been more than two decades since the first proposal of the concept of pseudo-haptics (Lecuyer et al., 2000), the adaptation of pseudo-haptic features into the design factors of buttons for modern VR is surprisingly lacking. A survey revealed that the design factors of buttons in VR such as sizes, positions, types, and shapes are not sufficiently addressed (Dube and Arif, 2019). Moreover, there have been very few comparative studies covering the fundamental factors of button design in VR (Bermejo et al., 2021). One recent preliminary study compared the planar (2D) and pseudo-haptic (3D) representations of UI widgets for menu interaction, and concluded the pseudo-haptic UI performs better in terms of workload, user experience, motion sickness, and immersion (Speicher et al., 2019). This demonstrates the potential of pseudo-haptic features for improving user experience thereby making other features worth studying in the domain of mid-air interaction in VR.

This study aims to investigate the effect of four pseudo-haptic features on the experience of the user touching the virtual button in mid-air. We conducted a user study to collect ratings on various subjective evaluation and user experience aspects when interacting with 16 buttons consist of different combinations of existence and nonexistence of four pseudo-haptic features. Then, we asked their preference towards each of the four pseudo-haptic features and the reasons behind their decisions.

## 2. Related Work

### *2.1. Free-Hand Mid-Air Interaction*

Mid-air interaction is a type of kinesthetic interaction (Fogtmann et al., 2008), denoting touchless and gesture-based interactions with displays or devices (Koutsabasis and Vogiatzidakis, 2019). Free-hand mid-air interaction possesses great potential in enabling natural UI in many user scenarios such as VR/AR applications, human-robot interactions, smart homes, and autonomous devices. In comparison with controller-based interfaces, users are free from holding a device and are allowed to engage with virtual objects directly via hand motions, promoting intuitiveness of the UI as spatial gestures are closely linked to the inherent manipulation skill of the human (Hinckley et al., 1994). Free-hand interfaces require tracking systems to detect mid-air hand gestures of the user as input. In virtue of extensive research on vision-based hand pose estimation (Ahmad et al., 2019; Erol et al., 2007; Supančič et al., 2018), modern commercial VR/AR HMDs are capable of detecting hands in real-time with embedded cameras.

Earlier works on free-hand mid-air interaction were explored across various modalities and application fields. Luo and Kenyon used scalable computing technologies to engage vision-based gesture interaction in a high-resolution tiled display setting (Luo and Kenyon, 2009). Hilliges et al. provided an easy way to handle digital contents in 3D by exploiting space above a standard interactive tabletop (Hilliges et al., 2009). Benko and Wilson combined free-hand pinch gestures with speech for the interaction of multiple users with a large curved display (Benko and Wilson, 2010). Nancel et al. developed a set of mid-air pan-and-zoom gestures to interact with graphics on a wall-sized display (Nancel et al., 2011). Song et al. studied designs of mid-air interaction in a free-hand setting for object manipulation in a 3D virtual space (Song et al., 2012).

Regarding direct interaction with floating virtual objects, several pioneering studies have explored using a Fresnel lens (Chan et al., 2010), a concave mirror (Butler et al., 2011), a see-through display (Hilliges et al., 2012), and a stereo projector (Benko et al., 2012), whereas later implementations occurred more in the context of VR HMD. Speicher et al. and Bermejo et al. tested design features of 3D UI for mid-air interaction in VR (Bermejo et al., 2021; Speicher et al., 2019), while Speicher et al. and Fashimpaur et al. emphasized improving user performance and experience for VR text entry (Fashimpaur et al., 2020; Speicher et al., 2018).

## 2.2. Pseudo-Haptics for Mid-Air Interaction

Most of the previous works on pseudo-haptics have simulated haptic properties by offering supplementary visual cues when the physical haptic cues were present by active devices or passive props (Abtahi et al., 2019; Abtahi and Follmer, 2018; Argelaguet et al., 2013; Ban et al., 2014; Dzidek et al., 2018; Hirano et al., 2011; Kimura and Nojima, 2012; Monnai et al., 2014; Punpongsanon et al., 2015, 2014; Samad et al., 2019; Sato et al., 2020; Ujitoko et al., 2015), thereby whether pseudo-haptic feedback could be made without a user's physical engagement with a tangible object remained uncertain. Interestingly, only a few studies attempted to create a haptic illusion in mid-air interaction when no external haptic inputs were given. Speicher et al. compared the experience of users for mid-air finger-based menu interaction with 2D and 3D UIs, and found 3D UI performed better (Speicher et al., 2019). Kawabe tested the stiffness of the virtual object with varied ratios of object deformation magnitude to the hand distance and found the deformation-to-distance ratio affected the perceived stiffness (Kawabe, 2020). Kang et al. revealed visual and auditory cues can influence perceived roughness on virtual mobile gadgets in mixed reality (Kang et al., 2021). These studies have shown the feasibility of pseudo-haptic feedback in mid-air interaction.

Haptic feedback supports interaction by providing users an important confirmation of a successful action. Therefore, visually augmented pseudo-haptic features can enhance haptic illusion or provide additional information to strengthen confirmation of the action to attain the same benefits. There is a large body of research investigating such features. One attempt was to use shadows as depth cues to enhance illusions when selecting or manipulating virtual objects. Herndon et al. visualized the spatial relationships between objects in 3D applications with shadows (Herndon et al., 1992), whereas Wanger investigated the effect of shadow sharpness and shape on the spatial relationships (Wanger, 1992). Chan et al. proposed the pseudo-shadow visual feedback which indicated the proximity of the finger on interacting with intangible displays, and found that feedback was helpful in improving user performance and satisfaction (Chan et al., 2010).

Another attempt was made in the design of UI widgets to adapt affordances of real-world objects. Planar 2D UIs were transformed into protruded pseudo-haptic 3D UIs following the look of real-world widgets (e.g. buttons, switches, sliders, and control knobs) in addition to the stereoscopic cues to further enhance the illusion like the user is interacting with tangible objects. The comparisons were made for the stereoscopic display (Zilch et al., 2014) and the VR HMD (Speicher et al., 2019), and reported superior performance of protruded 3D widgets in terms of usability and user experience ratings. Meanwhile, a recent study compared 2D and 3D buttons for the numeric keypad typing task, and their results implied the 3D buttons induced lesser button press depth when haptic cues are provided, while 2D buttons induced higher entry speed than the 3D counterparts (Bermejo et al., 2021).

One feature could be inspired from the field of gaming, where the game designers attempted to maximize the feeling of hit when the player attacks the enemy to improve enjoyment. This attempt bears resemblance to the pseudo-haptic approach in the sense that it creates a sensual illusion of collision with extra visual effects. Studies have examined the impact of various visual (e.g. particle, afterimage, vibration, and view adjustment) and auditory (e.g. shooting sound, explosion sound, and groan) effects in the 2D shooting game (Kim and Kim, 2004; Seo and Kim, 2010) and the 2D fighting game (Moon and Cho, 2012), and revealed significant improvements on the feeling of hit with such effects.

The other approach could be derived from the attempt to block the virtual hand from penetrating the virtual object, which eventually leads to the displacement between the real and virtual hands. This approach was first proposed by Lindeman et al. (2001), and was identified as evidence to prove visual dominance over proprioception (Burns et al., 2006) and haptic modality (Mensvoort et al., 2008). While extensive research has focused on utilizing hand displacement for the selection of the out-of-reach objects (Benda et al., 2020; Gonzalez and Follmer, 2019; Ogawa et al., 2020; Tian et al., 2020), limited efforts were made for pseudo-haptics. Pusch et al. applied hand displacement to provide haptic-like sensations by dynamically displacing the virtual hand in the simulation of force fields, and found most participants could feel force illusion that their hand was pushed inside the force field (Pusch et al., 2009). Rietzler et al. tested the pseudo-haptic feedback of the virtual hand not penetrating the static object, and reported a significant increase in immersion and enjoyment (Rietzler et al., 2018a). In this study, we adopt this feature to block the penetration of the finger in the context of button interaction in VR.

It is worthwhile to examine the pseudo-haptic features on their effect on the user experience for mid-air interaction in VR as the definitive conclusion cannot be made based on the prior works. In this study, we consider four aforementioned pseudo-haptic button design features: proximity feedback, protrusion, hit effect, and penetration blocking.

## 3. Methods

### 3.1. Participants

Twenty-one Korean young adults (13 males and 8 females) between the ages of 19 and 33 (M=23.8, SD=4.0) and with normal or corrected to normal vision participated in the experiment. All participants except one were right-handed. Eighteen participants have experienced a kind of headset-based VR applications before this experiment, but 16 of them used the VR headset no more than once a year, showing the majority of participants were light VR users. Six participants reported that they experienced mid-air interaction with the tracked virtual hand in VR. All participants gave consent for the experiment protocol approved by the University Institutional Review Board (IRB NO.: KH2021-009).

*3.2. Experimental Design*

Four pseudo-haptic features were investigated in this study (Figure 1). The first feature, proximity feedback (PF) provided visual feedback when the finger goes near the button by showing a circle that enlarges as the finger approaches the button. The use of pseudo-shadow was avoided to be independent of the direction of light of the virtual environment. The second feature, protrusion (PT) enabled the contact surface to be protruded then pushed into the base, mimicking the affordance of the real-world 3D button. The design of the button was adopted from the one proposed in a previous study (Speicher et al., 2019) and the example described in the guideline for Oculus developers (Facebook, 2021). The third feature, hit effect (HE) added an extra sparkling visual cue together with a small vibration of the button when triggered. Although shaking the screen induced the largest feeling of hit in a fighting game (Moon and Cho, 2012), it was not appropriate to be applied in VR since visually induced experiences of self-motion are known to cause motion sickness (Bonato et al., 2008). Therefore, two applicable effects: visual vibration and collision graphics were selected for this study. The last feature, penetration blocking (PB) prohibited penetration of the virtual hand which is shown to the user, although the real hand actually penetrated the button. Total 16 buttons with different combinations were created by enabling (O) or disabling (X) each of four features, and a 2 proximity feedback ($PF_O$, $PF_X$) × 2 protrusion ($PT_O$, $PT_X$) × 2 hit effect ($HE_O$, $HE_X$) × 2 penetration blocking ($PB_O$, $PB_X$) within-subject design was used to investigate the effect of those features (Figure 1).

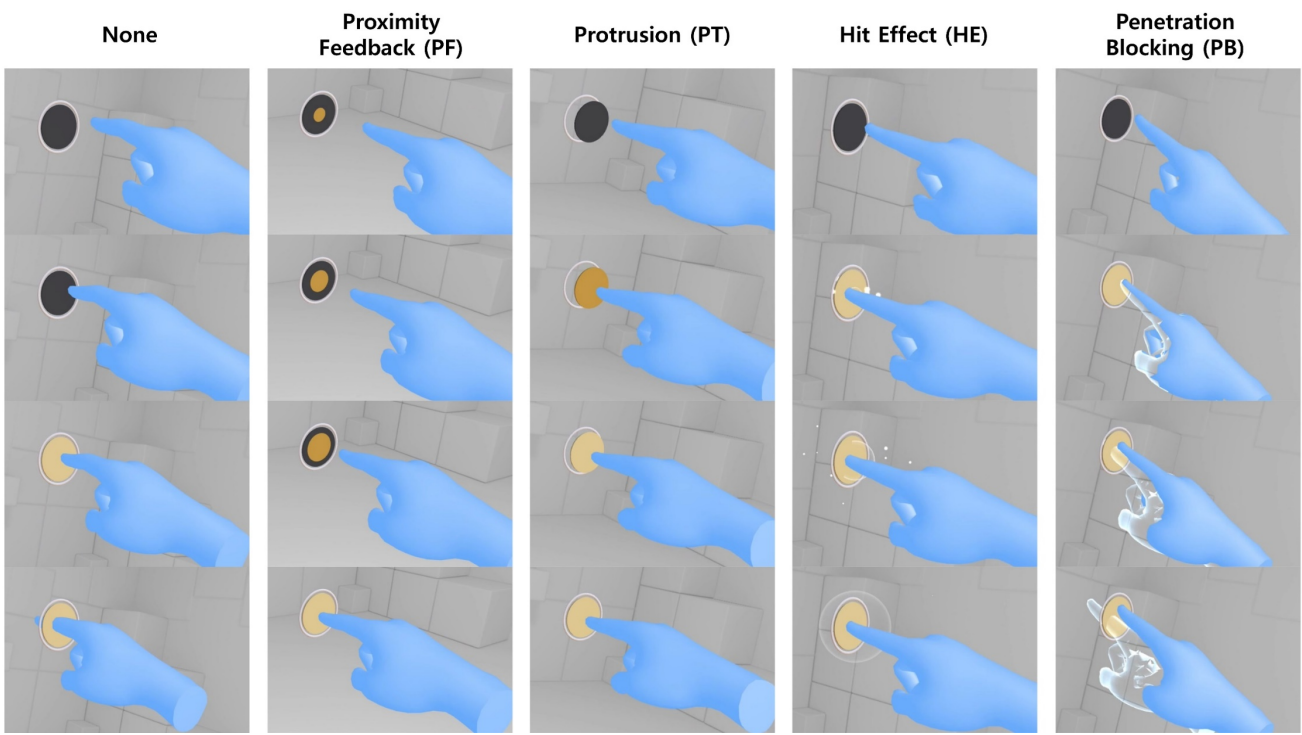

**Figure 1.** Four pseudo-haptic features investigated in this study. For penetration blocking, the location of the real hand, which is represented by a semi-transparent hand avatar, was invisible to the participants in the actual experiment. A video demonstration of 16 button combinations can be seen at the following link: https://vimeo.com/548802500.

*3.3. Experimental settings*

The participants experienced the virtual test environment with an Oculus Quest (resolution: 1440×1600 per eye; refresh rate: 72 Hz) VR headset, while the headset was connected to the PC through Oculus Link via a compatible USB 3.0 cable. The PC used to run the test environment was equipped with an Intel Core i7-7700 processor running at 3.6 GHz, 16 GB of RAM, and an NVIDIA GeForce GTX 1080 Ti GPU, running Windows 10. Hands were tracked in real-time by the four fisheye monochrome cameras embedded in the Oculus Quest headset with the help of an advanced hand detection network (Han et al., 2020). As hands were tracked by monochrome images, the background was covered by a black screen fence to prevent any potential deterioration of tracking performance (Figure 2-a). It is worth noting that although the virtual hand representation used in this study is not the most human-like hands, it does not severely violate the human-likeness thus does not significantly affect tactile experiences (Schwind et al., 2018). The screen fence was located 1.2 m away from participants, providing them enough space to freely interact with the virtual targets without any physical interruptions.

Inside the virtual test environment developed using Unity 2019.4.15f1, 16 circular buttons were placed in front of the participant in a 4×4 square grid (Figure 2-b). The center of the grid was located 0.2 m (scale unit in Unity) below the height of the VR headset when the participant was standing. The distance between two neighboring buttons placed in the same row or column was 0.1 m, and the diameter of the buttons was 5 cm. The participants were allowed to freely step back and forth to adjust the distance to the buttons to find the best setting for them to comfortably interact with the buttons. It is worthy to note that the location of buttons was determined to facilitate minimum ergonomic cost (Evangelista Belo et al., 2021) while taking the hand tracking range of the Oculus Quest headset into account to guarantee stable tracking performance at all times.

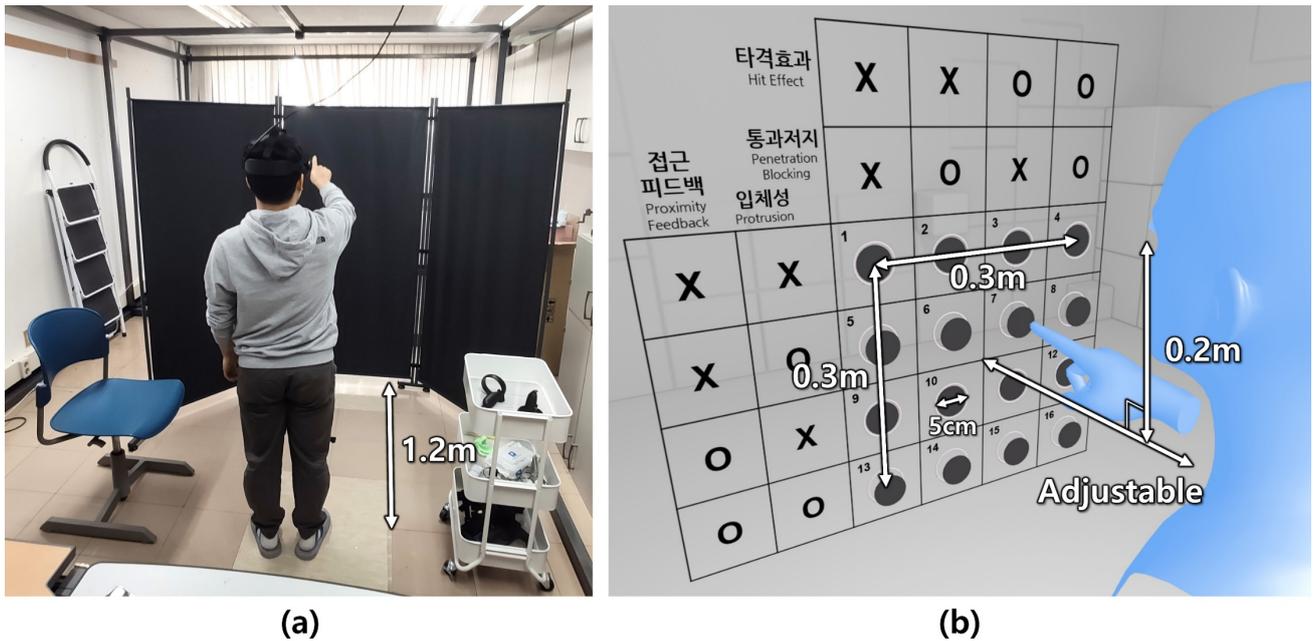

**Figure 2.** Experimental settings in (a) the real environment and (b) the virtual environment. The head avatar in the virtual environment was invisible to the participants in the actual experiment.

*3.4. Experimental Procedure*

First, participants filled in a pre-test questionnaire asking about their demographic information and prior experience with VR. After participants were explained about four pseudo-haptic features and the difference between 16 buttons, they were asked to press each of 16 buttons (Figure 2-b) with the index finger of their dominant hand twice following the pre-defined randomized order. Then, 11 subjective evaluation items about haptic illusion, embodiment, sense of reality, presence, spatiotemporal perception, and satisfaction aspects (Gonzalez-Franco and Peck, 2018; Pusch et al., 2009; Schwind et al., 2018; Usoh et al., 2000) were asked and the responses were collected through 7-point Likert scales (Table 1). In addition, a short version of the User Experience Questionnaire (Schrepp et al., 2017) was used to collect pragmatic and hedonic quality aspects of user experience (Table 1). After participants were asked to answer their preference on each feature and the corresponding reasons, they were encouraged to select a single most preferred button, although multiple selections were allowed in case they did not have a clear preference on certain features. During the whole questionnaire session, questionnaire items were presented right next to the buttons all the time, and participants verbally gave the rating in their preferred order while freely interacting with buttons as many times as they wanted.

**Table 1.** Items of the subjective evaluation and user experience questionnaires used in this study. Answers were given through 7-point Likert scales (Subjective evaluation: 1=strongly disagree, 7=strongly agree; User experience: -3=fully agree with the negative term, +3=fully agree with the positive term)

| | Aspect | No. | Item |
|---|---|---|---|
| **Subjective evaluation** | Haptic illusion | 1 | I felt the haptic sensation like touching the tangible object. |
| | | 2 | It seemed as if I felt the touch of the button in the location where I saw the virtual button touched. |
| | Embodiment | 3 | I felt as if the virtual hand was my own hand while pressing the button. |
| | | 4 | It felt like I could control the virtual hand as if it was my own hand while pressing the button. |
| | Sense of reality | 5 | My experiences in the virtual environment seemed consistent with my real-world experiences. |
| | | 6 | Interacting with this button felt like interacting with the real-world buttons. |
| | Presence | 7 | I had a sense of "being there" in the virtual space. |
| | | 8 | During the experience I often thought that I was really standing in the virtual space. |
| | Spatiotemporal perception | 9 | I could perceive the distance between my hand and the button well when pressing the button. |
| | | 10 | I could perceive the exact timing when the button will be pressed well. |
| | Satisfaction | 11 | I was satisfied with the overall experience of pressing the button. |
| **User experience** | Pragmatic quality | 1 | Obtrusive – Supportive |
| | | 2 | Complicated – Easy |
| | | 3 | Inefficient – Efficient |
| | | 4 | Clear – Confusing |
| | Hedonic quality | 5 | Boring – Exiting |
| | | 6 | Not interesting – Interesting |
| | | 7 | Conventional – Inventive |
| | | 8 | Usual – Leading edge |

*3.5. Data Analysis*

Ratings for items of the subjective evaluation and user experience questionnaires were collected and arranged for statistical analysis. For all aspects, mean ratings within each category were used for the analysis. No outliers in ratings were detected according to the Grubbs' outlier test. Analysis of variance (ANOVA) was conducted to check the statistical significance of each subjective evaluation item. Minitab 19 was used to conduct all statistical analyses at a significance level of 0.05. The variation from participants was blocked. The effect size in terms of eta-squared ($\eta^2$) was further calculated to check practical significance. A general rule on the magnitudes of the effect size with $\eta^2$ is as follows: small-$\eta^2 \sim 0.01$, medium-$\eta^2 \sim 0.06$ and large-$\eta^2 \sim 0.14$ (Cohen, 1988). For the selection of the most preferred button, the choices were normalized to keep the sum of frequency as 1 for each participant when multiple buttons were selected (e.g. 0.5 for selection of 2 buttons and 0.25 for 4 buttons).

## 4. Results

### 4.1. Subjective Ratings

#### 4.1.1. Main effects

Table 2 shows the ANOVA results on the ratings of subjective evaluation and user experience questionnaires. Significant main effects of PF, PT, HE, and PB were observed on the haptic illusion (PF: $F_{1,300}=13.79$, p<0.001; PT: $F_{1,300}=276.59$, p<0.001; HE: $F_{1,300}=241.27$, p<0.001; PB: $F_{1,300}=41.3$, p<0.001), embodiment (PF: $F_{1,300}=7.28$, p=0.007; PT: $F_{1,300}=118.64$, p<0.001; HE: $F_{1,300}=99.19$, p<0.001; PB: $F_{1,300}=5.59$, p=0.019), sense of reality (PF: $F_{1,300}=10.06$, p=0.002; PT: $F_{1,300}=147.00$, p<0.001; HE: $F_{1,300}=114.37$, p<0.001; PB: $F_{1,300}=45.93$, p<0.001), spatiotemporal perception (PF: $F_{1,300}=177.81$, p<0.001; PT: $F_{1,300}=172.27$, p<0.001; HE: $F_{1,300}=99.67$, p<0.001; PB: $F_{1,300}=21.88$, p<0.001), and satisfaction (PF: $F_{1,300}=10.97$, p=0.001; PT: $F_{1,300}=94.54$, p<0.001; HE: $F_{1,300}=161.68$, p<0.001; PB: $F_{1,300}=21.11$, p<0.001), whereas the main effect of HE was solely significant for presence ($F_{1,300}=37.84$, p<0.001). Ratings increased by enabling PF, PT, HE, and PB in all subjective evaluation aspects except presence. Among significant measures, the effect size differed between different pseudo-haptic features. For PF, large effect size ($\eta^2=0.159$) was found at spatiotemporal perception while the rest had small effect size ($\eta^2=0.009$-$0.014$). Medium to large effect size for PT ($\eta^2=0.125$-$0.229$) and HE ($\eta^2=0.089$-$0.200$) and small to medium effect size for PB ($\eta^2=0.007$-$0.047$) was found at all subjective evaluation ratings except presence. Figure 3 depicts mean ratings and differences for subjective evaluation on four pseudo-haptic features.

**Table 2.** P-values and $\eta^2$ from ANOVA results of effects of four pseudo-haptic features on the ratings of subjective evaluation and user experience questionnaires

| Questionnaire | Subjective evaluation | | | | | | | | | | | | User experience | | | |
|---|---|---|---|---|---|---|---|---|---|---|---|---|---|---|---|---|
| Aspect | Haptic illusion | | Embodiment | | Sense of reality | | Presence | | Spatiotemporal perception | | Satisfaction | | Pragmatic quality | | Hedonic quality | |
| Measure | p | $\eta^2$ | p | $\eta^2$ | p | $\eta^2$ | p | $\eta^2$ | p | $\eta^2$ | p | $\eta^2$ | p | $\eta^2$ | p | $\eta^2$ |
| **PF** | <.001*** | 0.011 | 0.007** | 0.009 | 0.002** | 0.010 | 0.184 | 0.004 | <.001*** | 0.159# | 0.001** | 0.014 | 0.025* | 0.011 | <.001*** | 0.109# |
| **PT** | <.001*** | 0.229# | <.001*** | 0.143# | <.001*** | 0.152# | 0.630 | 0.001 | <.001*** | 0.155# | <.001*** | 0.125# | 0.393 | 0.002 | <.001*** | 0.092# |
| **HE** | <.001*** | 0.200# | <.001*** | 0.120# | <.001*** | 0.118# | <.001*** | 0.082# | <.001*** | 0.089# | <.001*** | 0.213# | 0.007** | 0.016 | <.001*** | 0.403# |
| **PB** | <.001*** | 0.034 | 0.019* | 0.007 | <.001*** | 0.047 | 0.671 | <.001 | <.001*** | 0.020 | <.001*** | 0.028 | <.001*** | 0.033 | 0.012* | 0.006 |
| **PF×PT** | 0.077 | 0.003 | 0.444 | 0.001 | 1.000 | <.001 | 0.552 | 0.001 | 0.001** | 0.010 | 0.455 | 0.001 | 0.823 | <.001 | 0.012* | 0.006 |
| **PF×HE** | 0.821 | <.001 | 0.765 | <.001 | 0.950 | <.001 | 0.843 | <.001 | 1.000 | <.001 | 0.915 | <.001 | 0.742 | <.001 | 0.001** | 0.010 |
| **PF×PB** | 0.458 | <.001 | 0.868 | <.001 | 0.709 | <.001 | 0.843 | <.001 | 1.000 | <.001 | 0.455 | 0.001 | 0.906 | <.001 | 0.550 | <.001 |
| **PT×HE** | 0.001** | 0.010 | 0.714 | <.001 | 0.055 | 0.004 | 0.380 | 0.002 | 0.186 | 0.002 | 0.002** | 0.013 | 0.086 | 0.007 | 0.232 | 0.001 |
| **PT×PB** | 0.166 | 0.002 | 0.444 | 0.001 | 0.901 | <.001 | 0.977 | <.001 | 0.443 | 0.001 | 0.594 | <.001 | 0.408 | 0.002 | 0.816 | <.001 |
| **HE×PB** | 0.420 | 0.001 | 0.765 | <.001 | 0.154 | 0.002 | 0.184 | 0.004 | 0.126 | 0.002 | 0.594 | <.001 | 0.365 | 0.002 | 0.790 | <.001 |
| **PF×PT×HE** | 0.974 | <.001 | 0.303 | 0.001 | 0.804 | <.001 | 0.630 | 0.001 | 0.780 | <.001 | 0.337 | 0.001 | 0.554 | 0.001 | 0.790 | <.001 |
| **PF×PT×PB** | 0.540 | <.001 | 0.485 | 0.001 | 0.495 | <.001 | 0.843 | <.001 | 0.780 | <.001 | 0.749 | <.001 | 0.906 | <.001 | 0.618 | <.001 |
| **PF×HE×PB** | 0.233 | 0.001 | 0.816 | <.001 | 1.000 | <.001 | 0.755 | <.001 | 0.727 | <.001 | 0.749 | <.001 | 0.906 | <.001 | 0.947 | <.001 |
| **PT×HE×PB** | 0.583 | <.001 | 0.816 | <.001 | 0.901 | <.001 | 0.887 | <.001 | 0.889 | <.001 | 0.337 | 0.001 | 0.823 | <.001 | 0.973 | <.001 |
| **PF×PT×HE×PB** | 0.540 | <.001 | 0.572 | <.001 | 0.852 | <.001 | 0.515 | 0.001 | 0.727 | <.001 | 0.915 | <.001 | 0.927 | <.001 | 0.868 | <.001 |

\* PF=Proximity Feedback; PT=Protrusion; HE=Hit Effect; PB=Penetration Blocking; *p<0.05, **p<0.01, ***p<0.001, #$\eta^2$>0.06

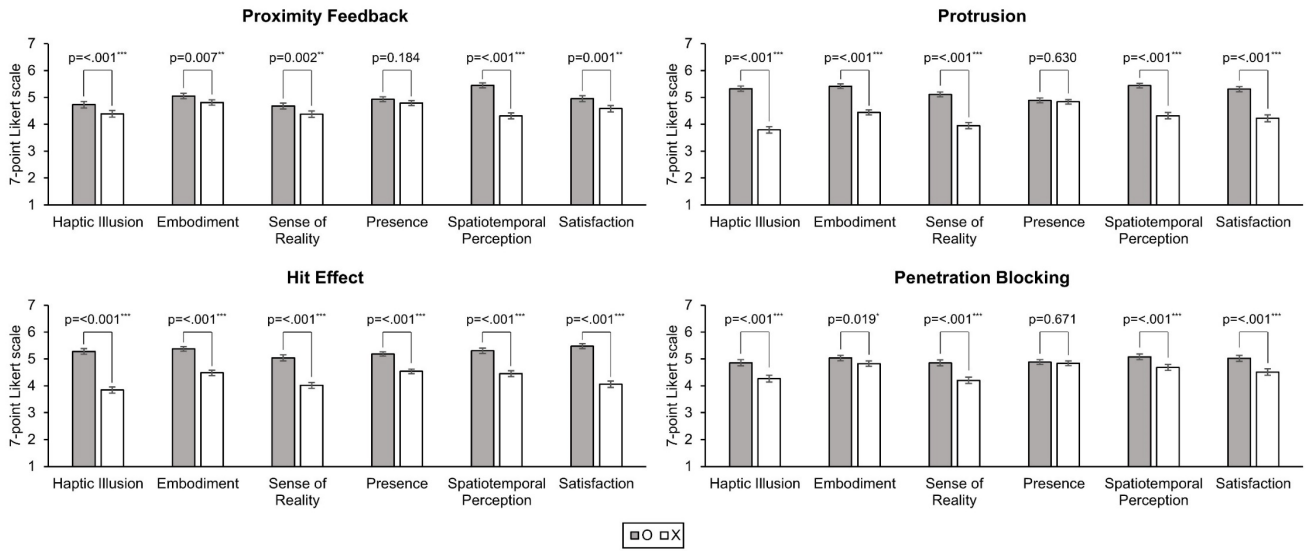

**Figure 3.** Mean ratings (+SE) and p-values from ANOVA results for subjective evaluation on four pseudo-haptic features. *Note:* *p<0.05, **p<0.01, ***p<0.001

Regarding the user experience, significant main effects of PF ($F_{1,300}$=5.06, p=0.025, $\eta^2$=0.011), HE ($F_{1,300}$=7.28, p=0.007, $\eta^2$=0.016), and PB ($F_{1,300}$=14.66, p<0.001, $\eta^2$=0.033) were observed on the pragmatic quality but with small effect sizes. HE and PB affected the pragmatic quality positively, whereas PF affected negatively. For hedonic quality, all features were found to have a significant positive effect: PF ($F_{1,300}$=124.27, p<0.001, $\eta^2$=0.109), PT ($F_{1,300}$=105.05, p<0.001, $\eta^2$=0.092), and HE ($F_{1,300}$=457.83, p<0.001, $\eta^2$=0.403) with medium to large effect sizes, and PB ($F_{1,300}$=6.4, p=0.012, $\eta^2$=0.006) with a small effect size. HE had the largest mean difference of rating (MD=1.91) and PB had the smallest difference (MD=0.23). PF, PT, and HE had medium to large effect sizes and PB had a small effect size. Figure 4 illustrates mean ratings and differences for user experience on four pseudo-haptic features.

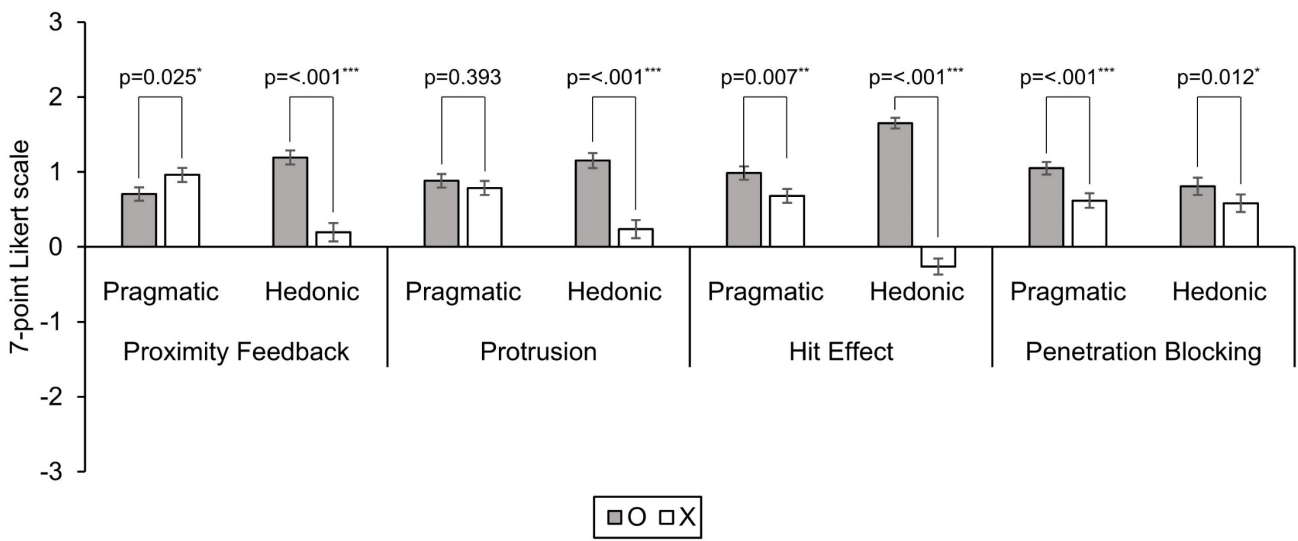

**Figure 4.** Mean ratings (+SE) and p-values from ANOVA results for user experience on four pseudo-haptic features. *Note:* *p<0.05, **p<0.01, ***p<0.001

*4.1.2. Interaction effects and correlation*

A few significant second-order interactions between PF, PT, and HE were found (Figure 5). More specifically, significant interaction effects were found in spatiotemporal perception ($F_{1,300}$=6.57, p=0.001, $\eta^2$=0.010) and hedonic quality ($F_{1,300}$=4.30, p=0.012, $\eta^2$=0.006) for PF × PT, hedonic quality ($F_{1,300}$=7.59, p=0.001, $\eta^2$=0.01) for PF × HE, and haptic illusion ($F_{1,300}$=, p=0.001, $\eta^2$=0.01) and satisfaction ($F_{1,300}$=, p=0.002, $\eta^2$=0.013) for PT × HE. The same tendency was found across all significant interaction effects where the degree of effect from these three features tends to decline when used together. However, it should be noted that the practical significances were considered negligible due to small effect sizes ($\eta^2$=0.006-0.013).

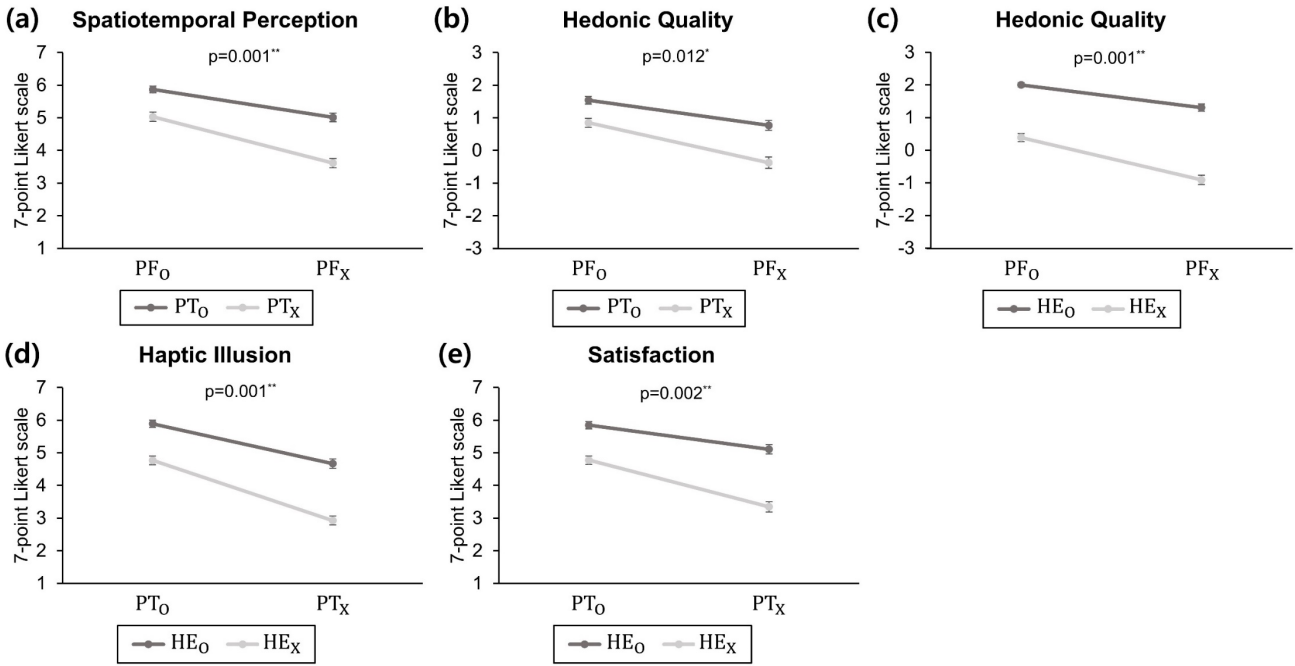

**Figure 5.** Mean ratings (+SE) and p-values from ANOVA results for subjective evaluation and user experience on ratings with significant interaction effects: (a-b) PF × PT, (c) PF × HE, and (d-e) PT × HE. *Note:* *p<0.05, **p<0.01, ***p<0.001

Table 3 shows the Spearman correlation matrix for collected eight different aspects of subjective ratings. The strongest correlation was found between haptic illusion and sense of reality ($\rho$=0.78), and the weakest correlation was found between presence and pragmatic quality ($\rho$=0.11). Aspects were moderately correlated to each other in general, although the correlation of presence and pragmatic quality with the rest tended to be weaker ($\rho$=0.11-0.44) compared to the correlation among the rest aspects ($\rho$=0.48-0.78). Satisfaction was correlated the most with haptic illusion ($\rho$=0.75) and the least with presence ($\rho$=0.28).

**Table 3.** Spearman correlation matrix for subjective ratings

|  | (1) | (2) | (3) | (4) | (5) | (6) | (7) | (8) |
|---|---|---|---|---|---|---|---|---|
| **(1) Haptic Illusion** | | | | | | | | |
| **(2) Embodiment** | 0.70 | | | | | | | |
| **(3) Sense of Reality** | 0.78 | 0.67 | | | | | | |
| **(4) Presence** | 0.30 | 0.40 | 0.22 | | | | | |
| **(5) Spatiotemporal Perception** | 0.61 | 0.48 | 0.67 | 0.19 | | | | |
| **(6) Satisfaction** | 0.75 | 0.61 | 0.69 | 0.28 | 0.60 | | | |
| **(7) Pragmatic Quality** | 0.32 | 0.32 | 0.39 | 0.11 | 0.23 | 0.44 | | |
| **(8) Hedonic Quality** | 0.61 | 0.51 | 0.51 | 0.31 | 0.61 | 0.62 | 0.17 | |

## *4.2. User Preference*

Figure 6-a shows the frequency of the preference on each of four pseudo-haptic features. HE and PT were most dominantly preferred (19 out of 21 participants, ~90%), followed by PF and PB with frequencies of 12 (57%) and 11 (52%), respectively. The number of participants who preferred to have the feature was larger compared to the number of participants who preferred to not have for all features. The number of participants who disliked the feature was the largest at PF with the frequency of 6 (29%), while the number of participants who did not have preference was the largest at PB with the frequency of 5 (24%). Figure 6-b further shows the frequency of the most preferred button among 16 buttons with different combinations of features. The button with all features enabled ($PF_O$, $PT_O$, $HE_O$, $PB_O$) was preferred the most with the normalized frequency of 8.75 (42%), followed by the button with all features but PF enabled ($PF_X$, $PT_O$, $HE_O$, $PB_O$) and the button with all features but PB enabled ($PF_O$, $PT_O$, $HE_O$, $PB_X$) with frequencies of 4.25 (20%) and 2.25 (11%), respectively. The top three buttons accounted for 73% of all selections.

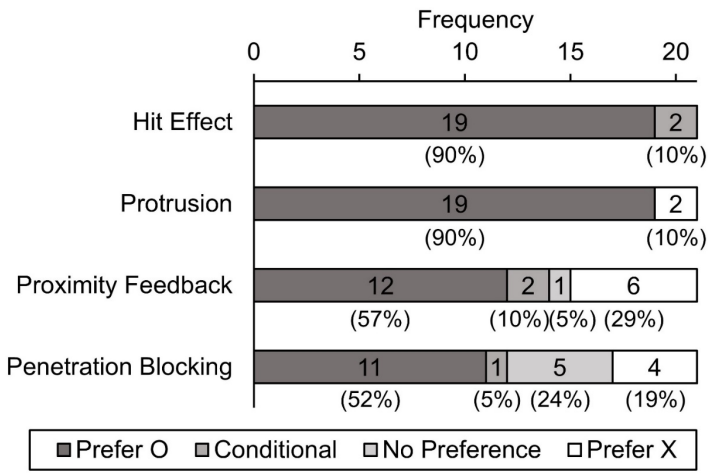

**Figure 6.** Frequency (percentage) of (a) the preference on each of four pseudo-haptic features and (b) the most preferred button

## 5. Discussion

### 5.1. Main Findings

PF majorly contributed to spatiotemporal perception as it should be, adhering to its original intention of aiding the perception of distance between the finger and the button by providing additional visual cues. Visual feedback about motion pattern and position coordinates significantly influences early and later stages of hand movement (Saunders and Knill, 2004), and enhances parallel processing of the visual and kinesthetic information about the ongoing hand movement (Sigrist et al., 2013). This could be especially useful in the immersive virtual environment where distances tend to be underestimated (Interrante et al., 2006; Knapp and Loomis, 2004). However, the negative effect of PF on the pragmatic quality indicating less value in traditional usability aspects (Schrepp et al., 2006) should be noted. Detailed analysis on individual questionnaire items revealed that participants felt the button with PF slightly more supportive ($F_{1,300}=7.90$, $p=0.005$, $\eta^2=0.012$) but moderately more complicated ($F_{1,300}=42.73$, $p<0.001$, $\eta^2=0.087$), which eventually led to lower ratings on the pragmatic quality. Chan et al. warned that continuous feedback could easily distract users as it provides extra visual cues during the entire interaction process but fails to inform the exact time the button is touched (Chan et al., 2010). The same result was observed in this study that 29% of participants disliked PF because it was distracting and continually gave undesirable pressure like they should press the button. Nevertheless, PF was favored by twice the number of participants (57%) and contributed to a slight improvement on other subjective aspects and a considerable improvement in the hedonic quality.

The positive effect of PT on user experience was expected as shown in several prior studies. Studies have shown 3D buttons were better than 2D buttons in terms of overall usability and memorability (Zilch et al., 2014); overall user experience, workload, motion sickness, and immersion (Speicher et al., 2019); naturalness, perceived performance, and preference (Dube and Arif, 2020). On top of that, this study newly revealed the improvements from the 3D button in haptic illusion, embodiment, sense of reality, and spatiotemporal perception. Participants found the interaction more interesting and feel more like pressing the real-world button when PT is enabled, which is in line with the study of Speicher et al. (2019) where participants found the affordance-oriented design of the 3D button more exciting and motivating. Participants in our study claimed PT also helped knowing when the button will be pressed, as the distance between the protruded surface and the base offered an extra visual cue. The cue acts similar to PF in a sense, but unlike PF, it relies on the stereoscopic depth cue of the button surface protruded towards the user. In addition, some participants thought it was a good match with other features; for example, three participants conditionally preferred to have other features only when the button is protruded (one participant for each of PF, HE, and PB).

By combining two effects: visual vibration and collision graphics, HE successfully emulated the feeling of hit and corresponding pseudo-haptic sensation. Visual vibration has been applied to the cursor to evoke pseudo-haptic surface roughness of the touch screen (Costes et al., 2019), and stylized visual effects have been explored in simulating haptic properties of touching real and virtual objects in AR (Mercado et al., 2020), but to the authors' knowledge, this is the first study investigated pseudo-haptics of these two visual effects in mid-air button interaction in VR. HE had a moderate to large impact on all subjective evaluation and user experience aspects except the pragmatic quality but especially, it induced the largest impact on satisfaction (MD=1.42) and hedonic quality (MD=1.91) among four investigated features. This result is reasonable considering this feature was inspired by special effects used in gaming, which put a huge emphasis on the entertainment and enjoyment of players. Interestingly, no participants disliked this feature. Participants claimed the hit effect not only made the interaction feel more interesting and tangible but also helped to know exactly when the button is pressed. Discrete feedback such as an instant visual transition or a sound effect is known to benefit interaction by clearly confirming a touch, yet provides little help before a touch (Chan et al., 2010). Therefore, the combination of PT and HE can be considered as a great harmony by offering advantages of both discrete and continuous feedback.

Notably, PT and HE were unconditionally preferred by 90% of participants, and buttons with both features enabled were chosen as the favorite by 79% of all choices (Figure 6-b).

The effect of PB was statistically significant but had a small practical impact on all subjective evaluation and user experience aspects except presence. 52% of participants who preferred PB claimed it enhanced haptic illusion and helped know when the button is pressed by checking the hand movement is blocked. They felt awkward, not realistic, and out of control when the hand penetrates the button because this is not what people with common sense will expect to happen in the real world. This becomes clear if we see the result that the sense of reality was influenced the most among all aspects, with the mean difference of 0.65 and eta-squared of 0.047. On the other hand, 19% of participants who did not like PB mentioned they felt their hand movement is restricted and awkward since it does not match with the movement of their real hand. This implies a caution is needed when applying this feature, as large offsets between the virtual and real hands over the threshold can place a big perceptional discrepancy between vision and proprioception (Burns et al., 2006). Although detection thresholds of hand redirection have been derived previously (horizontal/vertical offset: 4.5°, gain-based offset: 0.88-1.07; Zenner & Kruger, 2019), a further investigation to identify the offset threshold that can severely damage user experience in this specific case is needed, as even obvious perceivable offsets were accepted and preferred by VR users in some cases (Rietzler et al., 2018a, 2018b). 24% of participants who did not have a preference on this feature explained they could not identify an obvious difference caused by enabling PB. As participants were not requested to press the button beyond the base when they interact with the button, buttons with and without PB could be perceived as indistinguishable for some participants who always stopped their finger at the base before penetration.

It is worth mentioning the popular reasons for preference in all features were: (1) I felt an illusion of a reactive force or a touch (haptic illusion), (2) I felt like pressing the real-world button (sense of reality), (3) I easily knew the finger distance and the moment the button is pressed (spatiotemporal perception), and (4) I felt more interesting (hedonic quality). Haptic illusion ($\rho=0.75$), sense of reality ($\rho=0.69$), spatiotemporal perception ($\rho=0.60$), and hedonic quality ($\rho=0.62$) showed moderate to strong correlation with satisfaction. Our results provide a piece of empirical evidence that pseudo-haptic button design features could elevate such perception-related items thereby improve the user experience of free-hand mid-air interaction in VR.

According to the results, all four proposed pseudo-haptic button design features either majorly or minorly contributed to improving user experience. However, it is doubtful whether more features will always lead to a better user experience. Although the button with all features enabled was chosen by the largest number of participants as their favorite, some results proved otherwise. Some participants pointed out a certain feature (especially PF) feels distracting when too many features are used together, thereby either disliked or conditionally liked corresponding features. For instance, two participants preferred PF and HE under the condition of enabling only one of the two. Their concerns were also reflected in the ratings as significant interaction effects which indicated the decrease of feature impact at combined usage among PF, PT, and HE.

*5.2. Design Implications*

Our research findings implied that all four investigated pseudo-haptic features (PF, PT, HE, and PB) could benefit users in terms of user experience. VR developers can consider adapting proposed features into the design of 3D UI to improve perceived haptic illusion, embodiment, sense of reality, spatiotemporal perception, satisfaction, and hedonic quality of users. However, it should be noted that the effect of PF and PB were relatively minor along with a nonnegligible portion of disfavor compared to PT and HE. Therefore, an option to disable specific features could be beneficial for certain users, as individuals comprehend and interpret the haptic experience in different ways hence might want to tailor effects for their preferences (Schneider et al., 2017). The proposed features can be applied to general 3D UIs with interactive buttons in any type of VR application to provide a more satisfactory and enjoyable experience to users. Except for PB which relies on a virtual representation of the real hand, we expect the remaining three features: PF, PT, and HE can be applied to 3D UIs in other application domains (e.g. AR) to merit users.

### *5.3. Limitations and Future Work*

This study has some limitations. First of all, since this study mainly focused on investigating the effect of pseudo-haptic features on user experience, the effect on task performance was not considered. A recent study from Bermejo et al. reported that the users with 2D buttons unintuitively achieved faster key entry speed than the 3D buttons in the numeric keypad typing task (Bermejo et al., 2021), showing the possibility that button protrusion may result in worse performance in certain task scenarios, although Dube and Arif reported contradictory results that 3D keys yielded lower error rate compared to 2D keys in the VR text entry task (Dube and Arif, 2020). Future work can further verify the effect of proposed features on the task performance for interacting with 3D UI or virtual keyboard where mid-air touch interaction with virtual buttons can frequently occur. Second, this study attempted to investigate the pure effect of visual cues hence excluded the auditory cues which were proven to be effective in addition to visual cues for multi-modal conditions (Bermejo et al., 2021; Chan et al., 2010; Kobayashi et al., 2016; Lecuyer et al., 2001). The addition of auditory cues with proper sensory integration strategy to further improve the user experience of mid-air interaction in VR can be investigated in future work. Third, the hit effect consisted of two visual components: a visual vibration effect and a collision graphics effect, so it is unclear whether the positive impact mainly came from the particle effect or the vibration effect. These two effects could not be investigated separately to keep the number of factors and experimental conditions manageable in this study, but a more concrete investigation on individual effects deserves consideration in the future.

## 6. Conclusion

We conducted a user study with 21 young subjects to investigate the effect of four pseudo-haptic button design features (proximity feedback, protrusion, hit effect, and penetration blocking) on user experience for free-hand mid-air button interaction in VR. All investigated features significantly and positively affected user experience in terms of haptic illusion, embodiment, sense of reality, spatiotemporal perception, satisfaction, and hedonic quality. In addition, protrusion and hit effect were more beneficial in comparison with the other two features. It is suggested to use the four proposed pseudo-haptic features for designing 3D UI to make users feel more satisfied and entertained, but caution is needed when using proximity feedback together with other features. The findings of this study may serve as a useful input for VR developers and UI designers to create better interactive buttons in 3D UI.

### Acknowledgments

This work was supported by the Basic Science Research Program through the National Research Foundation of Korea funded by the Ministry of Science, ICT and Future Planning (NRF-2020R1F1A1048510).

**References**

Abtahi, P., Follmer, S., 2018. Visuo-haptic illusions for improving the perceived performance of shape displays, in: Conference on Human Factors in Computing Systems - Proceedings. ACM, New York, NY, USA, pp. 1–13. https://doi.org/10.1145/3173574.3173724

Abtahi, P., Landry, B., Yang, J. (Junrui), Pavone, M., Follmer, S., Landay, J.A., 2019. Beyond The Force: Using Quadcopters to Appropriate Objects and the Environment for Haptics in Virtual Reality, in: Proceedings of the 2019 CHI Conference on Human Factors in Computing Systems. ACM, New York, NY, USA, pp. 1–13. https://doi.org/10.1145/3290605.3300589

Ahmad, A., Migniot, C., Dipanda, A., 2019. Hand pose estimation and tracking in real and virtual interaction:A review. Image Vis. Comput. 89, 35–49. https://doi.org/10.1016/j.imavis.2019.06.003

Arafsha, F., Zhang, L., Dong, H., Saddik, A. El, 2015. Contactless haptic feedback: state of the art, in: 2015 IEEE International Symposium on Haptic, Audio and Visual Environments and Games (HAVE). IEEE, pp. 1–6. https://doi.org/10.1109/HAVE.2015.7359447

Argelaguet, F., Jauregui, D.A.G., Marchal, M., LeCuyer, A., 2013. Elastic images: Perceiving local elasticity of images through a novel pseudo-haptic deformation effect. ACM Trans. Appl. Percept. 10. https://doi.org/10.1145/2501599

Azmandian, M., Hancock, M., Benko, H., Ofek, E., Wilson, A.D., 2016. Haptic Retargeting: Dynamic Repurposing of Passive Haptics for Enhanced Virtual Reality Experiences, in: Proceedings of the 2016 CHI Conference on Human Factors in Computing Systems. ACM, New York, NY, USA, pp. 1968–1979. https://doi.org/10.1145/2858036.2858226

Ban, Y., Narumi, T., Tanikawa, T., Hirose, M., 2014. Controlling perceived stiffness of pinched objects using visual feedback of hand deformation, in: IEEE Haptics Symposium, HAPTICS. pp. 557–562. https://doi.org/10.1109/HAPTICS.2014.6775516

Benda, B., Esmaeili, S., Ragan, E.D., 2020. Determining Detection Thresholds for Fixed Positional Offsets for Virtual Hand Remapping in Virtual Reality, in: 2020 IEEE International Symposium on Mixed and Augmented Reality (ISMAR). IEEE, pp. 269–278. https://doi.org/10.1109/ISMAR50242.2020.00050

Benko, H., Jota, R., Wilson, A., 2012. MirageTable: freehand interaction on a projected augmented reality tabletop, in: Proceedings of the 2012 ACM Annual Conference on Human Factors in Computing Systems - CHI '12. ACM Press, New York, New York, USA, p. 199. https://doi.org/10.1145/2207676.2207704

Benko, H., Wilson, A.D., 2010. Multi-point interactions with immersive omnidirectional visualizations in a dome, in: ACM International Conference on Interactive Tabletops and Surfaces - ITS '10. ACM Press, New York, New York, USA, p. 19. https://doi.org/10.1145/1936652.1936657

Benko, H., Wilson, A.D., Baudisch, P., 2006. Precise selection techniques for multi-touch screens, in: Proceedings of the SIGCHI Conference on Human Factors in Computing Systems - CHI '06. ACM Press, New York, New York, USA, p. 1263. https://doi.org/10.1145/1124772.1124963

Bermejo, C., Lee, L.H., Chojecki, P., Przewozny, D., Hui, P., 2021. Exploring Button Designs for Mid-air Interaction in Virtual Reality: A Hexa-metric Evaluation of Key Representations and Multi-modal Cues. Proc. ACM Human-Computer Interact. 5, 1–26. https://doi.org/10.1145/3457141

Blake, J., Gurocak, H.B., 2009. Haptic Glove With MR Brakes for Virtual Reality. IEEE/ASME Trans. Mechatronics 14, 606–615. https://doi.org/10.1109/TMECH.2008.2010934

Bonato, F., Bubka, A., Palmisano, S., Phillip, D., Moreno, G., 2008. Vection Change Exacerbates Simulator Sickness in Virtual Environments. Presence Teleoperators Virtual Environ. 17, 283–292. https://doi.org/10.1162/pres.17.3.283

Bouzit, M., Burdea, G., Popescu, G., Boian, R., 2002. The Rutgers Master II - New design force-feedback glove. IEEE/ASME Trans. Mechatronics 7, 256–263. https://doi.org/10.1109/TMECH.2002.1011262

Bowman, D.A., McMahan, R.P., Ragan, E.D., 2012. Questioning naturalism in 3D user interfaces. Commun. ACM 55, 78–88. https://doi.org/10.1145/2330667.2330687

Bullion, C., Gurocak, H., 2009. Haptic Glove with MR Brakes for Distributed Finger Force Feedback. Presence Teleoperators Virtual Environ. 18, 421–433. https://doi.org/10.1162/pres.18.6.421

Burns, E., Razzaque, S., Panter, A.T., Whitton, M.C., McCallus, M.R., Brooks, F.P., 2006. The hand is more easily fooled than the eye: Users are more sensitive to visual interpenetration than to visual-proprioceptive discrepancy. Presence Teleoperators Virtual Environ. 15, 1–15. https://doi.org/10.1162/pres.2006.15.1.1

Butler, A., Hilliges, O., Izadi, S., Hodges, S., Molyneaux, D., Kim, D., Kong, D., 2011. Vermeer: Direct interaction with a 360° viewable 3D display, in: UIST'11 - Proceedings of the 24th Annual ACM Symposium on User Interface Software and Technology. ACM Press, New York, New York, USA, pp. 569–576. https://doi.org/10.1145/2047196.2047271

Caggianese, G., Gallo, Luigi, Neroni, P., 2019. The vive controllers vs. leap motion for interactions in virtual environments: A comparative evaluation, in: De Pietro, G., Gallo, L, Howlett, R., Jain, L., Vlacic, L. (Eds.), Smart Innovation, Systems and Technologies. Springer, Cham, pp. 24–33. https://doi.org/10.1007/978-3-319-92231-7_3

Carter, T., Seah, S.A., Long, B., Drinkwater, B., Subramanian, S., 2013. UltraHaptics: Multi-point mid-air haptic feedback for touch surfaces, in: UIST 2013 - Proceedings of the 26th Annual ACM Symposium on User Interface Software and Technology. ACM, New York, NY, USA, pp. 505–514. https://doi.org/10.1145/2501988.2502018

Chan, L.-W., Kao, H.-S., Chen, M.Y., Lee, M.-S., Hsu, J., Hung, Y.-P., 2010. Touching the void, in: Proceedings of the 28th International Conference on Human Factors in Computing Systems - CHI '10. ACM Press, New York, New York, USA, p. 2625. https://doi.org/10.1145/1753326.1753725

Chattopadhyay, D., 2016. Understanding Interaction Mechanics In Touchless Target Selection. Indiana University. https://doi.org/10.7912/C27G73

Cohen, J., 1988. Statistical power analysis for the behavioural science, 2nd ed, Lawrence Erlbaum Associates. Hillsdale, NJ.

Costes, A., Argelaguet, F., Danieau, F., Guillotel, P., Lécuyer, A., 2019. Touchy : A Visual Approach for Simulating Haptic Effects on Touchscreens. Front. ICT 6. https://doi.org/10.3389/fict.2019.00001

Dube, T.J., Arif, A.S., 2020. Impact of Key Shape and Dimension on Text Entry in Virtual Reality, in: Extended Abstracts of the 2020 CHI Conference on Human Factors in Computing Systems. ACM, New York, NY, USA, pp. 1–10. https://doi.org/10.1145/3334480.3382882

Dube, T.J., Arif, A.S., 2019. Text Entry in Virtual Reality: A Comprehensive Review of the Literature, in: M., K. (Ed.), Human-Computer Interaction. Recognition and Interaction Technologies. HCII 2019. Lecture Notes in Computer Science, Vol 11567. Springer, Cham, pp. 419–437. https://doi.org/10.1007/978-3-030-22643-5_33

Dudley, J., Benko, H., Wigdor, D., Kristensson, P.O., 2019. Performance Envelopes of Virtual Keyboard Text Input Strategies in Virtual Reality, in: 2019 IEEE International Symposium on Mixed and Augmented Reality (ISMAR). IEEE, pp. 289–300. https://doi.org/10.1109/ISMAR.2019.00027

Dzidek, B., Frier, W., Harwood, A., Hayden, R., 2018. Design and Evaluation of Mid-Air Haptic Interactions in an Augmented Reality Environment, in: Prattichizzo, D., Shinoda, H., Tan, H., Ruffaldi, E., Frisoli, A. (Eds.), Haptics: Science, Technology, and Applications. EuroHaptics 2018. Lecture Notes in Computer Science, Vol 10894. Springer, Cham, pp. 489–499. https://doi.org/10.1007/978-3-319-93399-3_42

Erol, A., Bebis, G., Nicolescu, M., Boyle, R.D., Twombly, X., 2007. Vision-based hand pose estimation: A review. Comput. Vis. Image Underst. 108, 52–73. https://doi.org/10.1016/j.cviu.2006.10.012

Evangelista Belo, J.M., Feit, A.M., Feuchtner, T., Grønbæk, K., 2021. XRgonomics: Facilitating the Creation of Ergonomic 3D Interfaces, in: Proceedings of the 2021 CHI Conference on Human Factors in Computing Systems. ACM, New York, NY, USA, pp. 1–11. https://doi.org/10.1145/3411764.3445349

Facebook, 2021. User Interface Components [WWW Document]. Oculus Dev. URL https://developer.oculus.com/learn/hands-design-ui/

Faeth, A., Harding, C., 2014. Emergent effects in multimodal feedback from virtual buttons. ACM Trans. Comput. Interact. 21, 1–23. https://doi.org/10.1145/2535923

Fashimpaur, J., Kin, K., Longest, M., 2020. PinchType: Text Entry for Virtual and Augmented Reality Using Comfortable Thumb to Fingertip Pinches, in: Extended Abstracts of the 2020 CHI Conference on Human Factors in Computing Systems. ACM, New York, NY, USA, pp. 1–7. https://doi.org/10.1145/3334480.3382888

Fogtmann, M.H., Fritsch, J., Kortbek, K.J., 2008. Kinesthetic interaction: Revealing the bodily potential in interaction

design, in: Proceedings of the 20th Australasian Conference on Computer-Human Interaction Designing for Habitus and Habitat - OZCHI '08. ACM Press, New York, New York, USA, p. 89. https://doi.org/10.1145/1517744.1517770

Freeman, E., Vo, D.-B., Brewster, S., 2019. HaptiGlow: Helping Users Position their Hands for Better Mid-Air Gestures and Ultrasound Haptic Feedback, in: 2019 IEEE World Haptics Conference (WHC). IEEE, pp. 289–294. https://doi.org/10.1109/WHC.2019.8816092

Gabardi, M., Solazzi, M., Leonardis, D., Frisoli, A., 2016. A new wearable fingertip haptic interface for the rendering of virtual shapes and surface features, in: 2016 IEEE Haptics Symposium (HAPTICS). IEEE, pp. 140–146. https://doi.org/10.1109/HAPTICS.2016.7463168

Gonzalez-Franco, M., Peck, T.C., 2018. Avatar Embodiment. Towards a Standardized Questionnaire. Front. Robot. AI 5. https://doi.org/10.3389/frobt.2018.00074

Gonzalez, E.J., Follmer, S., 2019. Investigating the Detection of Bimanual Haptic Retargeting in Virtual Reality, in: 25th ACM Symposium on Virtual Reality Software and Technology. ACM, New York, NY, USA, pp. 1–5. https://doi.org/10.1145/3359996.3364248

Hachisu, T., Cirio, G., Marchal, M., Lecuyer, A., Kajimoto, H., 2011. Pseudo-haptic feedback augmented with visual and tactile vibrations, in: 2011 IEEE International Symposium on VR Innovation. IEEE, pp. 327–328. https://doi.org/10.1109/ISVRI.2011.5759662

Han, S., Liu, B., Cabezas, R., Twigg, C.D., Zhang, P., Petkau, J., Yu, T.-H., Tai, C.-J., Akbay, M., Wang, Z., Nitzan, A., Dong, G., Ye, Y., Tao, L., Wan, C., Wang, R., 2020. MEgATrack: monochrome egocentric articulated hand-tracking for virtual reality. ACM Trans. Graph. 39. https://doi.org/10.1145/3386569.3392452

Herndon, K.P., Van Dam, A., Gleicher, M., 1994. The challenges of 3D interaction. ACM SIGCHI Bull. https://doi.org/10.1145/191642.191652

Herndon, K.P., Zeleznik, R.C., Robbins, D.C., Conner, D.B., Snibbe, S.S., van Dam, A., 1992. Interactive shadows, in: Proceedings of the 5th Annual ACM Symposium on User Interface Software and Technology - UIST '92. ACM Press, New York, New York, USA, pp. 1–6. https://doi.org/10.1145/142621.142622

Hilliges, O., Izadi, S., Wilson, A.D., Hodges, S., Garcia-Mendoza, A., Butz, A., 2009. Interactions in the air: adding further depth to interactive tabletops, in: Proceedings of the 22nd Annual ACM Symposium on User Interface Software and Technology - UIST '09. ACM Press, New York, New York, USA, p. 139. https://doi.org/10.1145/1622176.1622203

Hilliges, O., Kim, D., Izadi, S., Weiss, M., Wilson, A., 2012. HoloDesk: direct 3d interactions with a situated see-through display, in: Proceedings of the 2012 ACM Annual Conference on Human Factors in Computing Systems - CHI '12. ACM Press, New York, New York, USA, p. 2421. https://doi.org/10.1145/2207676.2208405

Hinckley, K., Pausch, R., Goble, J.C., Kassell, N.F., 1994. A survey of design issues in spatial input, in: Proceedings of the 7th Annual ACM Symposium on User Interface Software and Technology - UIST '94. ACM Press, New York, New York, USA, pp. 213–222. https://doi.org/10.1145/192426.192501

Hirano, Y., Kimura, A., Shibata, F., Tamura, H., 2011. Psychophysical influence of mixed-reality visual stimulation on sense of hardness, in: Proceedings - IEEE Virtual Reality. pp. 51–54. https://doi.org/10.1109/VR.2011.5759436

Hoggan, E., Brewster, S.A., Johnston, J., 2008. Investigating the effectiveness of tactile feedback for mobile touchscreens, in: Proceeding of the Twenty-Sixth Annual CHI Conference on Human Factors in Computing Systems - CHI '08. ACM Press, New York, New York, USA, p. 1573. https://doi.org/10.1145/1357054.1357300

Hwang, I., Son, H., Kim, J.R., 2017. AirPiano: Enhancing music playing experience in virtual reality with mid-air haptic feedback, in: 2017 IEEE World Haptics Conference (WHC). IEEE, pp. 213–218. https://doi.org/10.1109/WHC.2017.7989903

Interrante, V., Ries, B., Anderson, L., 2006. Distance Perception in Immersive Virtual Environments, Revisited, in: IEEE Virtual Reality Conference (VR 2006). IEEE, pp. 3–10. https://doi.org/10.1109/VR.2006.52

Janlert, L.E., 2014. The ubiquitous button. Interactions 21, 26–33. https://doi.org/10.1145/2592234

Joyce, R.D., Robinson, S., 2017. Passive Haptics to Enhance Virtual Reality Simulations, in: AIAA Modeling and Simulation Technologies Conference. American Institute of Aeronautics and Astronautics, Reston, Virginia.

https://doi.org/10.2514/6.2017-1313

Kang, N., Sah, Y.J., Lee, S., 2021. Effects of visual and auditory cues on haptic illusions for active and passive touches in mixed reality. Int. J. Hum. Comput. Stud. 150, 102613. https://doi.org/10.1016/j.ijhcs.2021.102613

Kawabe, T., 2020. Mid-Air Action Contributes to Pseudo-Haptic Stiffness Effects. IEEE Trans. Haptics 13, 18–24. https://doi.org/10.1109/TOH.2019.2961883

Kim, H., Kim, M., Lee, W., 2016. HapThimble: A Wearable Haptic Device towards Usable Virtual Touch Screen, in: Proceedings of the 2016 CHI Conference on Human Factors in Computing Systems. ACM, New York, NY, USA, pp. 3694–3705. https://doi.org/10.1145/2858036.2858196

Kim, N., Kim, T., 2004. A Study on the Advance Effect of Hitting Sense in Shooting Game -Center for "the Beetlewing" and '1945 Plus'-. J. Korea Multimed. Soc. 7, 223–230.

Kimura, T., Nojima, T., 2012. Pseudo-haptic feedback on softness induced by grasping motion. Lect. Notes Comput. Sci. (including Subser. Lect. Notes Artif. Intell. Lect. Notes Bioinformatics) 7283 LNCS, 202–205. https://doi.org/10.1007/978-3-642-31404-9_36

Knapp, J.M., Loomis, J.M., 2004. Limited Field of View of Head-Mounted Displays Is Not the Cause of Distance Underestimation in Virtual Environments. Presence Teleoperators Virtual Environ. 13, 572–577. https://doi.org/10.1162/1054746042545238

Kobayashi, A., Aoki, R., Kitagawa, N., Kimura, T., Takashima, Y., Yamada, T., 2016. Towards enhancing force-input interaction by visual-auditory feedback as an introduction of first use, in: Lecture Notes in Computer Science (Including Subseries Lecture Notes in Artificial Intelligence and Lecture Notes in Bioinformatics). Springer-Verlag, Berlin, Heidelberg, pp. 180–191. https://doi.org/10.1007/978-3-319-39516-6_17

Koutsabasis, P., Vogiatzidakis, P., 2019. Empirical Research in Mid-Air Interaction: A Systematic Review. Int. J. Human–Computer Interact. 35, 1747–1768. https://doi.org/10.1080/10447318.2019.1572352

LaViola Jr., J.J., Kruijff, E., McMahan, R.P., Bowman, D., Poupyrev, I.P., 2017. 3D user interfaces: theory and practice. 3D User Interfaces Theory Pract.

Lecuyer, A., Burkhardt, J.-M., Coquillart, S., Coiffet, P., 2001. "Boundary of illusion": an experiment of sensory integration with a pseudo-haptic system, in: Proceedings IEEE Virtual Reality 2001. IEEE Comput. Soc, pp. 115–122. https://doi.org/10.1109/VR.2001.913777

Lecuyer, A., Coquillart, S., Kheddar, A., Richard, P., Coiffet, P., 2000. Pseudo-haptic feedback: can isometric input devices simulate force feedback?, in: Proceedings IEEE Virtual Reality 2000 (Cat. No.00CB37048). IEEE Comput. Soc, pp. 83–90. https://doi.org/10.1109/VR.2000.840369

Lee, J., An, S.-G., Kim, Y., Bae, S.-H., 2018. Projective Windows: Bringing Windows in Space to the Fingertip. Ext. Abstr. 2018 CHI Conf. Hum. Factors Comput. Syst. D412:1--D412:1. https://doi.org/10.1145/3170427.3186524

Lee, L.-H., Hui, P., 2018. Interaction Methods for Smart Glasses: A Survey. IEEE Access 6, 28712–28732. https://doi.org/10.1109/ACCESS.2018.2831081

Lindeman, R.W., Sibert, J.L., Templeman, J.N., 2001. The effect of 3D widget representation and simulated surface constraints on interaction in virtual environments, in: Proceedings IEEE Virtual Reality 2001. IEEE Comput. Soc, pp. 141–148. https://doi.org/10.1109/VR.2001.913780

Luo, X., Kenyon, R.V., 2009. Scalable Vision-based Gesture Interaction for Cluster-driven High Resolution Display Systems, in: 2009 IEEE Virtual Reality Conference. IEEE, pp. 231–232. https://doi.org/10.1109/VR.2009.4811030

Mensvoort, K. van, Hermes, D.J., Montfort, M. van, 2008. Usability of optically simulated haptic feedback. Int. J. Hum. Comput. Stud. 66, 438–451. https://doi.org/10.1016/j.ijhcs.2007.12.004

Mercado, V., Normand, J.-M., Lecuyer, A., 2020. "Kapow!": Augmenting Contacts with Real and Virtual Objects Using Stylized Visual Effects, in: 2020 IEEE International Symposium on Mixed and Augmented Reality Adjunct (ISMAR-Adjunct). IEEE, pp. 116–117. https://doi.org/10.1109/ISMAR-Adjunct51615.2020.00043

Monnai, Y., Hasegawa, K., Fujiwara, M., Yoshino, K., Inoue, S., Shinoda, H., 2014. HaptoMime: Mid-Air Haptic Interaction with a Floating Virtual Screen, in: Proceedings of the 27th Annual ACM Symposium on User Interface Software and Technology. ACM, New York, NY, USA, pp. 663–667. https://doi.org/10.1145/2642918.2647407

Moon, S.-J., Cho, H.-J., 2012. A Study on Enhancing Efficiency for Feeling-of-Hit in Games. J. Korea Game Soc. 12, 3–14. https://doi.org/10.7583/JKGS.2012.12.2.3

Nancel, M., Wagner, J., Pietriga, E., Chapuis, O., Mackay, W., 2011. Mid-air pan-and-zoom on wall-sized displays, in: Proceedings of the 2011 Annual Conference on Human Factors in Computing Systems - CHI '11. ACM Press, New York, New York, USA, p. 177. https://doi.org/10.1145/1978942.1978969

Ogawa, N., Narumi, T., Hirose, M., 2020. Effect of Avatar Appearance on Detection Thresholds for Remapped Hand Movements. IEEE Trans. Vis. Comput. Graph. 1–1. https://doi.org/10.1109/TVCG.2020.2964758

Punpongsanon, P., Iwai, D., Sato, K., 2015. SoftAR: Visually manipulating haptic softness perception in spatial augmented reality. IEEE Trans. Vis. Comput. Graph. 21, 1279–1288. https://doi.org/10.1109/TVCG.2015.2459792

Punpongsanon, P., Iwai, D., Sato, K., 2014. A preliminary study on altering surface softness perception using augmented color and deformation, in: ISMAR 2014 - IEEE International Symposium on Mixed and Augmented Reality - Science and Technology 2014, Proceedings. pp. 301–302. https://doi.org/10.1109/ISMAR.2014.6948460

Pusch, A., Martin, O., Coquillart, S., 2009. HEMP-hand-displacement-based pseudo-haptics: A study of a force field application and a behavioural analysis. Int. J. Hum. Comput. Stud. 67, 256–268. https://doi.org/10.1016/j.ijhcs.2008.09.015

Rietzler, M., Geiselhart, F., Frommel, J., Rukzio, E., 2018a. Conveying the Perception of Kinesthetic Feedback in Virtual Reality using State-of-the-Art Hardware, in: Proceedings of the 2018 CHI Conference on Human Factors in Computing Systems. ACM, New York, NY, USA, pp. 1–13. https://doi.org/10.1145/3173574.3174034

Rietzler, M., Geiselhart, F., Gugenheimer, J., Rukzio, E., 2018b. Breaking the Tracking: Enabling Weight Perception using Perceivable Tracking Offsets, in: Proceedings of the 2018 CHI Conference on Human Factors in Computing Systems. ACM, New York, NY, USA, pp. 1–12. https://doi.org/10.1145/3173574.3173702

Samad, M., Gatti, E., Hermes, A., Benko, H., Parise, C., 2019. Pseudo-Haptic Weight: Changing the Perceived Weight of Virtual Objects By Manipulating Control-Display Ratio, in: Proceedings of the 2019 CHI Conference on Human Factors in Computing Systems. ACM, New York, NY, USA, pp. 1–13. https://doi.org/10.1145/3290605.3300550

Sato, Y., Hiraki, T., Tanabe, N., Matsukura, H., Iwai, D., Sato, K., 2020. Modifying Texture Perception With Pseudo-Haptic Feedback for a Projected Virtual Hand Interface. IEEE Access 8, 120473–120488. https://doi.org/10.1109/ACCESS.2020.3006440

Saunders, J.A., Knill, D.C., 2004. Visual Feedback Control of Hand Movements. J. Neurosci. 24, 3223–3234. https://doi.org/10.1523/JNEUROSCI.4319-03.2004

Schneider, O., MacLean, K., Swindells, C., Booth, K., 2017. Haptic experience design: What hapticians do and where they need help. Int. J. Hum. Comput. Stud. 107, 5–21. https://doi.org/10.1016/j.ijhcs.2017.04.004

Schrepp, M., Held, T., Laugwitz, B., 2006. The influence of hedonic quality on the attractiveness of user interfaces of business management software. Interact. Comput. 18, 1055–1069. https://doi.org/10.1016/j.intcom.2006.01.002

Schrepp, M., Hinderks, A., Thomaschewski, J., 2017. Design and Evaluation of a Short Version of the User Experience Questionnaire (UEQ-S). Int. J. Interact. Multimed. Artif. Intell. 4, 103. https://doi.org/10.9781/ijimai.2017.09.001

Schwind, V., Lin, L., Di Luca, M., Jörg, S., Hillis, J., 2018. Touch with foreign hands: The effect of virtual hand appearance on visual-haptic integration, in: Proceedings - SAP 2018: ACM Symposium on Applied Perception. ACM, New York, NY, USA, pp. 1–8. https://doi.org/10.1145/3225153.3225158

Seo, J., Kim, N., 2010. Empirical Analysis of the Feeling of Shooting in 2D Shooting Games. J. Korea Soc. Comput. Inf. 15, 75–81.

Shakeri, G., Williamson, J.H., Brewster, S., 2018. May the Force Be with You: Ultrasound Haptic Feedback for Mid-Air Gesture Interaction in Cars, in: Proceedings of the 10th International Conference on Automotive User Interfaces and Interactive Vehicular Applications. ACM, New York, NY, USA, pp. 1–10. https://doi.org/10.1145/3239060.3239081

Shneiderman, B., 1997. Direct manipulation for comprehensible, predictable and controllable user interfaces, in: International Conference on Intelligent User Interfaces, Proceedings IUI. pp. 33–39. https://doi.org/10.1145/238218.238281

Sigrist, R., Rauter, G., Riener, R., Wolf, P., 2013. Augmented visual, auditory, haptic, and multimodal feedback in motor learning: A review. Psychon. Bull. Rev. 20, 21–53. https://doi.org/10.3758/s13423-012-0333-8

Song, P., Goh, W.B., Hutama, W., Fu, C.-W., Liu, X., 2012. A handle bar metaphor for virtual object manipulation with mid-air interaction, in: Conference on Human Factors in Computing Systems - Proceedings. pp. 1297–1306. https://doi.org/10.1145/2207676.2208585

Speicher, M., Ehrlich, J., Gentile, V., Degraen, D., Sorce, S., Krüger, A., 2019. Pseudo-haptic Controls for Mid-air Finger-based Menu Interaction, in: Extended Abstracts of the 2019 CHI Conference on Human Factors in Computing Systems. ACM, New York, NY, USA, pp. 1–6. https://doi.org/10.1145/3290607.3312927

Speicher, M., Feit, A.M., Ziegler, P., Krüger, A., 2018. Selection-based Text Entry in Virtual Reality, in: Proceedings of the 2018 CHI Conference on Human Factors in Computing Systems - CHI '18. ACM Press, New York, New York, USA, pp. 1–13. https://doi.org/10.1145/3173574.3174221

Strandholt, P.L., Dogaru, O.A., Nilsson, N.C., Nordahl, R., Serafin, S., 2020. Knock on Wood: Combining Redirected Touching and Physical Props for Tool-Based Interaction in Virtual Reality, in: Proceedings of the 2020 CHI Conference on Human Factors in Computing Systems. ACM, New York, NY, USA, pp. 1–13. https://doi.org/10.1145/3313831.3376303

Supančič, J.S., Rogez, G., Yang, Y., Shotton, J., Ramanan, D., 2018. Depth-Based Hand Pose Estimation: Methods, Data, and Challenges. Int. J. Comput. Vis. 126, 1180–1198. https://doi.org/10.1007/s11263-018-1081-7

Tian, Y., Bai, Y., Zhao, S., Fu, C.W., Yang, T., Heng, P.A., 2020. Virtually-Extended Proprioception: Providing Spatial Reference in VR through an Appended Virtual Limb, in: Conference on Human Factors in Computing Systems - Proceedings. ACM, New York, NY, USA, pp. 1–12. https://doi.org/10.1145/3313831.3376557

Ujitoko, Y., Tanikawa, T., Ban, Y., Hirota, K., Narumi, T., Hirose, M., 2015. Yubi-toko: Finger walking in snowy scene using pseudo-haptic technique on touchpad, in: SIGGRAPH Asia 2015 Emerging Technologies, SA 2015. https://doi.org/10.1145/2818466.2818491

Ungureanu, D., Bogo, F., Galliani, S., Sama, P., Duan, X., Meekhof, C., Stühmer, J., Cashman, T.J., Tekin, B., Schönberger, J.L., Olszta, P., Pollefeys, M., 2020. HoloLens 2 Research Mode as a Tool for Computer Vision Research.

Usoh, M., Catena, E., Arman, S., Slater, M., 2000. Using presence questionnaires in reality. Presence Teleoperators Virtual Environ. 9, 497–503. https://doi.org/10.1162/105474600566989

van Dam, A., 1997. Post-WIMP user interfaces. Commun. ACM 40, 63–67. https://doi.org/10.1145/253671.253708

Wanger, L., 1992. The effect of shadow quality on the perception of spatial relationships in computer generated imagery, in: Proceedings of the Symposium on Interactive 3D Graphics. ACM Press, New York, New York, USA, pp. 39–42. https://doi.org/10.1145/147156.147161

Zenner, A., Kruger, A., 2019. Estimating Detection Thresholds for Desktop-Scale Hand Redirection in Virtual Reality, in: 2019 IEEE Conference on Virtual Reality and 3D User Interfaces (VR). IEEE, pp. 47–55. https://doi.org/10.1109/VR.2019.8798143

Zhao, X., Niikura, T., Komuro, T., 2014. Evaluation of Visuo-haptic Feedback in a 3D Touch Panel Interface, in: Proceedings of the Ninth ACM International Conference on Interactive Tabletops and Surfaces - ITS '14. ACM Press, New York, New York, USA, pp. 299–304. https://doi.org/10.1145/2669485.2669536

Zilch, D., Bruder, G., Steinicke, F., 2014. Comparison of 2D and 3D GUI Widgets for Stereoscopic Multitouch Setups. J. Virtual Real. Broadcast. 11. https://doi.org/10.20385/1860-2037/11.2014.7